# Non-Volatile Superconductivity in an Insulating Copper Oxide Induced via Ionic Liquid Gating

*Xinjian Wei, Hao-bo Li, Qinghua Zhang, Dong Li, Mingyang Qin, Wei Hu, Ge He, Qing Huan, Li Yu, Qihong Chen, Jun Miao, Jie Yuan, Beiyi Zhu, A. Kusmartseva, F. V. Kusmartsev, Alejandro V. Silhanek, Tao Xiang, Weiqiang Yu, Yuan Lin, Lin Gu, Pu Yu, and Kui Jin\**

X. Wei, Dr. Q. Zhang, D. Li, M. Qin, W. Hu, Dr. G. He, Prof. Q. Huan, Dr. L. Yu, Dr. J. Yuan, Dr. B. Zhu, Prof. T. Xiang, Prof. L. Gu, Prof. K. Jin
Beijing National Laboratory for Condensed Matter Physics, Institute of Physics, Chinese Academy of Sciences, Beijing 100190, China
E-mail: kuijin@iphy.ac.cn
X. Wei, D. Li, M. Qin, W. Hu, Dr. G. He, Prof. T. Xiang, Prof. L. Gu, Prof. K. Jin
School of Physical Sciences, University of Chinese Academy of Sciences, Beijing 100049, China
Dr. Q. Chen
Device Physics of Complex Materials, Zernike Institute for Advanced Materials, University of Groningen, Groningen 9747 AG, The Netherlands
Prof. T. Xiang, Prof. L. Gu, Prof. P. Yu, Prof. K. Jin
Collaborative Innovation Center of Quantum Matter, Beijing 100190, China
Dr. J. Yuan, Prof. K. Jin
Songshan Lake Materials Laboratory, Dongguan, Guangdong 523808, China
Dr. H.-B. Li, Prof. P. Yu
State Key Laboratory of Low Dimensional Quantum Physics and Department of Physics, Tsinghua University, Beijing 100084, China
Prof. J. Miao
State Key Laboratory for Advanced Metals and Materials, School of Materials Science and Engineering, University of Science and Technology Beijing, Beijing 100083, China
Prof. A. Kusmartseva, Prof. F. V. Kusmartsev
Department of Physics, Loughborough University, LE11 3TU Loughborough, United Kingdom
Prof. A. V. Silhanek
Experimental Physics of Nanostructured Materials, Q-MAT, CESAM, Université de Liège B-4000 Sart Tilman, Belgium
Prof. W. Yu
Department of Physics, Renmin University of China, Beijing 100872, China
Prof. Y. Lin
State Key Laboratory of Electronic Thin Films and Integrated Devices & Center for Information in Medicine, University of Electronic Science and Technology of China, Chengdu 610054, China

**Manipulating the superconducting states of high-$T_c$ cuprate superconductors in an efficient and reliable way is of great importance for their applications in next-generation electronics. Traditional methods are mostly based on a trial-and-error method that is** **difficult to implement and time consuming. Here, employing ionic liquid gating, a selective control of volatile and non-volatile superconductivity is achieved in pristine insulating $Pr_2CuO_{4\pm\delta}$ film, based on two distinct mechanisms: 1) with positive electric fields, the film can be reversibly switched between non-superconducting and superconducting states, attributed to the carrier doping effect. 2) The film becomes more resistive by applying negative bias voltage up to -4 V, but strikingly, a non-volatile superconductivity is achieved once the gate voltage is removed. Such a persistent superconducting state represents a novel phenomenon in copper oxides, resulting from the doping healing of oxygen vacancies in copper-oxygen planes as unraveled by high-resolution scanning transmission electron microscope and *in-situ* x-ray diffraction experiments. The effective manipulation and mastering of volatile/non-volatile superconductivity in the same parent cuprate opens the door to more functionalities for superconducting electronics, as well as supplies flexible samples for investigating the nature of quantum phase transitions in high-$T_c$ superconductors**.

Copper oxide (cuprate) superconductors continue to be considered technologically attractive materials mainly due to their high transition temperature (high-$T_c$). The highest $T_c$ is commonly achieved by doping charge carriers, either cation substitutions or oxygen variations, into the parent compound known as Mott insulator.[1, 2] Although high-$T_c$ cuprate superconductors have been discovered for more than three decades, their applications are in a relatively stagnated state. One of the major reasons is that the optimal superconductivity and physical properties of cuprates are very sensitive to subtle changes of chemical composition, which is hard to control during the synthesis procedures.[3, 4] At the early stage, much effort was made to optimize the superconducting properties via a trial-and-error method.[5, 6] Recently, a complex two-step annealing method has been proposed to tune the oxygen content and induce superconductivity in the insulating parent cuprates.[7] However, such annealing is performed at several hundred degrees Celsius, well above the bearable temperature of semiconductor devices, which limits the integration of superconductors with other materials. Hence, exploring a low-temperature and cost-efficient method to manipulate the superconductivity will greatly help further promote the practical applications of high-$T_c$ cuprates.

Recently, ionic liquid gating (ILG) technique has emerged as an attractive method to trigger phase transitions by tuning the charge carriers in a large scale of $10^{14}$-$10^{15}$ cm$^{-2}$ due to the electrostatic field effect.[8-13] For example, ILG can turn ZrNCl[14] or MoS$_2$[15] into superconductors, and activate the superconducting state in La$_{2-x}$Sr$_x$CuO$_4$.[16] Such process is reversible as the superconductivity vanishes after removing the gate voltage. Very recently, non-volatile superconductivity was induced in Fe-based compounds by protonation yet the origin of this superconductivity and the role of protons need to be clarified.[17] Besides field effect and protonation, ILG could also change oxygen content via electrochemical interactions and drive the sample into a new electronic phase. This happens, for example, in SrCoO$_{2.5}$,[18] VO$_2$,[19] and SrTiO$_3$.[20] Nevertheless, it has never been reported that controlling oxygen content by ILG could turn an insulator into a superconductor. Therefore, it is tempting to manipulate the superconductivity of cuprates by modulating the oxygen content with ILG, which can be *in-situ* controlled and therefore save people from huge amounts of trial-and-error work.

For this purpose, a prototypical parent cuprate Pr$_2$CuO$_{4\pm\delta}$ (PCO) was chosen as our model system. The PCO consists of alternatively stacking of fluorite-like rare earth layers and square-planar CuO$_2$ layers, which is the parent compound of electron-doped (Pr,Ce)$_2$CuO$_4$. High-quality single-crystal PCO thin films were successfully prepared using the polymer assisted deposition method.[21] By carefully tuning oxygen content via annealing process, a superconducting dome can be obtained as seen in the blue region of **Figure 1**a.[22] Therefore, this system provides an ideal platform to study high-$T_c$ superconductivity from these non-superconducting PCO thin films with the ILG method. We find that positive gate voltage (PGV) can drive the insulating PCO thin films into superconducting one, forming a dome-like superconducting phase in the temperature-voltage (*T-V*) phase diagram (**SC I** in Figure 1(a)), mimicking the superconducting dome in phase diagram of *T*-Ce substitution.[3] Such superconducting state disappears as the PGV is withdrawn, consistent with the previous reports on $La_{2-x}Ce_xCuO_4$ and $Pr_{2-x}Ce_xCuO_4$.[23, 24] This phenomenon is commonly interpreted by electron injection into the sample through the PGV process (see Figure 1(b)).

Strikingly, although the sample becomes more resistive with negative gate voltage (NGV) to -4 V, a persistent superconducting state (i.e. non-volatile superconductivity) emerges right after the NGV is removed. A corresponding superconducting dome feature via this NGV operation is presented in the *T-V* phase diagram (**SC II** in Figure 1(a)) along with the **SC I** dome. To our best knowledge, this novel persistent superconducting phenomenon induced in insulating copper oxide via NGV process has never been reported before, and it provides a new and more practical strategy of manipulating superconductivity in high-$T_c$ cuprates via ILG technique. Combined with high-resolution scanning transmission electron microscope (HR-STEM) and *in-situ* x-ray diffraction (XRD) measurements, we thereby propose here a model that the repairing of oxygen vacancies in the $CuO_2$ plane under negative voltage gating (see Figure 1c) can be a rather good candidate to understand such new phenomenon.

**Figure 2**a shows the temperature dependence of resistance measured under various PGVs for sample S1. The bias voltage of ILG was set at 250 K (see the Experimental Section for more details). With increasing PGV, the normal state resistance gradually decreases and the sample becomes superconducting when PGV reaches 2.1 V. However, this superconductivity disappears once PGV is switched off. To detect the variation of carrier density with gating, we carried out Hall resistivity measurements. As can be seen in Figure 2b, the Hall resistivity $\rho_{xy}$

gradually increases with increasing PGV, similar to the effect of substituting $Pr^{3+}$ by $Ce^{4+}$ from under-doped regime to the optimally doping area.[25, 26] These suggest that this PGV-induced superconductivity (**SC I**) is caused by electron doping, which is further supported by the gating experiments on a pristine superconducting sample named as S2. As shown in Figure 2c, PGV can push the superconducting state of sample S2 into a metallic state, mimicking the superconductor-to-metal transition from optimally doped to heavily overdoped regime in Ce-doped materials.[1]

In the negative bias gating process, as seen in Figure 2d, the resistance of sample S1 shows an indiscernible change in the voltage range from 0 to -3 V (i.e. resistance curves almost overlap) whereas a remarkable increase occurs at -4 V. Surprisingly, a superconducting state emerges after the gate voltage is turned off. Correspondingly, the Hall resistivity barely changes in the voltage range from 0 to -3 V but drops substantially at -4 V, as illustrated in Figure 2e. When the gate voltage is removed, there is an abrupt sign change of Hall resistivity from negative to positive (see Figure 2e). A similar phenomenon has been observed in $Pr_{2-x}Ce_xCuO_4$ with increasing the Ce-doped concentration from underdoped to overdoped.[25, 26] However, a simple scenario based on electrostatic carrier doping cannot account for the nature of non-volatile superconductivity because the cycle of 0 → -4 → 0 V is unlikely to cause electron doping effect. In other words, the superconducting phase SC II arises from a novel and peculiar effect, which is dramatically distinct from the PGV-induced volatile superconductivity in our experiment and non-volatile superconductivity caused by PGV-driven proton injection in Fe-based compounds [17]. In order to further verify that the **SC II** phase appearing after a cycle of NGV is indeed non-volatile, we did the measurements for another non-superconducting sample S3. As shown in Figure 2f, the first cycle of PGV to 2.5 V cannot drive the sample to a superconducting state. However, superconductivity emerges after the sample undergoes a NGV process at -4 V. A natural question one may ask is that what happens to the samples as the NGV reaches a threshold?

Previous studies demonstrate that NGV above a threshold could trigger an electrochemical reaction in some oxides, e.g. $SrCoO_{2.5}$,[18] $YBa_2Cu_3O_{7-x}$,[27] and $La_2CuO_4$.[28] That is, oxygen ions are introduced into the samples. In order to get an explicit picture of the non-volatile superconductivity we observed, *in-situ* XRD measurements were performed on an insulating PCO thin film at room temperature (see the Experimental Section for details). As shown in **Figure 3**, the almost unchanged full width at half maxima (FWHM, see Figure 3b) and angle (Figure 3c) of (006) Bragg peak suggest minute crystallographic or structure factor modification along the *c*-axis in ILG process. This also indicates that no extra oxygen ions are introduced to the apical site, otherwise the *c*-axis lattice constant should increase.[29, 30] Notably, the amplitude of (006) Bragg peak is significantly enhanced as the NGV is set up to -2.5 V, highlighted by the red lines in Figure 3a and red symbols in Figure 3d. Such enhancement is retained after the gate bias is switched off, implying irreversible modification of the XRD form factor. In other words, the atomic electron density distribution and crystal structure change within the building block $CuO_2$ plane induced by negative voltage operation. Here the room-temperature threshold bias voltage for NGV operation is about -2.5 V for the *in-situ* XRD measurements. Such threshold voltage for electrochemical interaction is strongly *T*-dependent [31, 32], which becomes -4 V at 250 K as mentioned above.

In order to identify the origin of such crystallographic and/or electronic related changes in the compound, we compare the HR-STEM and EELS results between superconducting (SC) and non-superconducting (NSC) samples. **Figures 4**a and 4b are the HR-STEM images with single atom resolution along the [110] zone axis in the angular bright field mode. Figures 4d and 4e display the corresponding line profiles of $CuO_2$ layers in the atomic columns highlighted in Figures 4a and 4b, respectively. The spatial distribution of oxygen with atomic resolution can be reflected by the amplitude fluctuations of the lines connected by the shallow valleys (inverted triangles), which are more undulating for the NSC sample than for the SC one. This difference demonstrates more oxygen vacancies existing in the NSC sample.

Furthermore, the oxygen K-edge spectra of electron energy loss spectroscopy (EELS) from both NSC and SC samples are also presented in Figure 4f. It is clear that the superconducting sample has a much more pronounced leading edge peak structures in the EELS spectra, mainly related to the oxygen *p*-orbital hybridizing with the Cu *d*-orbital upper Hubbard branch from the $CuO_2$ planes (Figure 4f)[33]. Especially the major peak features around absorption threshold between 530 and 560 eV in the SC sample show clear spectroscopic distinction with respect to the NSC sample. To be more specific, the peak feature around 535 eV presents clear red shift to its counterpart in NSC spectra, and peak around 545 eV pops out more pronouncedly from spectral background in the superconducting sample. All these suggest clear electronic structure modification associated directly with oxygen ions in the $CuO_2$ planes induced by the NGV process.

It is known that unavoidable water inside the ionic liquid can be decomposed into negatively charged $O^{2-}$ and positively charged $H^+$ through electrolysis if sufficiently high gate voltage is applied.[18] Under NGV, anions accumulate on the surface of the sample. Since the oxygen vacancies in $CuO_2$ planes actually act as positively charged centers, they will attract negatively charged oxygen ions $O^{2-}$ once the electrostatic potential can overcome the crystal lattice energy. Therefore, it is reasonable to speculate that the -4 V bias plays a role in repairing the oxygen vacancies electrochemically within the $CuO_2$ planes, and consequently alters some local electron density distribution, which is consistent with the variation of peak amplitude in the *in-situ* XRD measurements.

Finally, we discuss possible mechanisms for the non-volatile superconductivity. Considerable oxygen vacancies in $CuO_2$ planes act as potential barriers, breaking Cooper pairs and preventing the pristine samples from showing superconductivity.[34] It is known that there exists a charge transfer gap between the upper Hubbard band (Cu $d_{x^2-y^2}$ orbit) and the O 2*p* band in cuprates.[35] The charge transfer gap can be smeared out by chemical doping [36] or field effect doping[37], due to the enhanced Coulomb screening effect and weakened on-site

Coulomb repulsion.[38] It should be pointed out that the non-superconducting PCO samples used here for ILG are not ideal Mott insulators, but rather slightly Mott doped insulators. This means that charge transfer gap in our non-superconducting samples is soft. At the bias voltage -4 V, one can suspect that on one hand, the electrochemical reaction process heals oxygen vacancies in the $CuO_2$ plane and thus reduces the pair-break scattering; on the other hand, the charge transfer gap is substantially enhanced, causing a sharp decrease of Hall resistivity and the absence of superconductivity. Once the gate voltage returns from -4 to 0 V, the charge transfer gap turns to close, thereby allowing the emergence of superconductivity. Since electrons and holes come from upper Hubbard band and O $2p$ band respectively, the mixture of these two bands results in the coexistence of two-type carriers. Such speculation can explain the non-linear magnetic field dependence of Hall resistivity as the NGV is removed, but requires that hole-type carriers dominate the transport for a positive Hall resistivity (Figure 2e). Another possibility is that besides the replenishing of oxygen vacancies, some oxygen ions are introduced to the interstitial sites, meanwhile bringing hole-type carriers into the film. Withdrawing the electric field, the interstitial oxygen ions accumulate near the film surface, resulting in a remarkable hole doping effect at the surface and thereby allowing the emergence of superconductivity. This scenario is more charming since hole-doped superconductor is achieved from the prototype of electron-doped cuprate superconductor.[39] In any case, the repair of $CuO_2$ plane is the key to realize non-volatile superconductivity.

In summary, we find that two different states of superconductivity can be effectively induced in insulating ternary copper oxide $Pr_2CuO_{4\pm\delta}$ by employing the ILG technique. One is volatile, resulting from electron carrier doping with PGV. The other is nonvolatile, due to a replenishing of oxygen vacancies in the $CuO_2$ planes with NGV. Our findings provide a new paradigm for inducing and manipulating superconductivity in copper oxide superconductors, one which could lead to a more detailed experimental exploring of the fundamental physics of the high-$T_c$ superconductivity as well as a route to optimize the superconductor for practical applications. Furthermore, an inexpensive and easy-to-implement method to control oxygen vacancies with several volts of gate bias is technically extendable to other oxygen-sensitive functional materials.

**Experimental Section**

*Film growth*: PCO thin films were grown on (00l)-oriented SrTiO$_3$ substrates by polymer assisted deposition method. The as-grown samples were fired at 850 ℃ in a tubular furnace with oxygen pressure about 200 Pa for crystallization. Then these samples were annealed at 400–600 ℃ under oxygen pressure of 1 Pa. By adjusting the annealing temperature and time, samples with various $T_c$ could be obtained. The film thickness is about 60 nm.

*In-situ Transport and XRD Measurements*: PCO samples were placed in a small quartz-glass cup and covered entirely by the ionic liquid, N-diethyl-N-methyl-N(2-methoxyethyl)-ammonium bis-(triuoromethylsulfonyl)imide (DEME-TFSI). We carried out electrical transport measurements of various PCO thin films in the physical property measurement system equipped with a Keithley 2400 source meter to apply the gate voltages at 250 K. The voltage dwell time is 20 min. In addition, we performed *in-situ* XRD measurements in air and at room temperature, by a high-resolution diffractometer (Smartlab, Rigaku) with monochromatic Cu K$_{α1}$ radiation (λ=1.5406 Å). The gate voltages vary from -2.5 to 2.5 V in steps of 0.5 V. Since the gating device shields the low-angle X-rays, the 2θ range was set from 41 to 49 degrees.

*HR-STEM experiments*: The crystalline structures of SC and NSC samples were investigated by the HR-STEM (ARM-200CF, from JEOL) equipped with double spherical aberration (Cs) correctors. The PCO specimens were observed along the [110] zone axis, which were prepared by focused ion beam techniques. The experiments were performed at 200 keV and the collection angle of angular bright-field images was 11-23 mrad.

**Acknowledgements**

We thank Q. Li, X. Zhang and F. C. Zhang for fruitful discussions. This work was supported by the National Key Basic Research Program of China (2015CB921000, 2017YFA0302900, 2017YFA0303003 and 2018YFB0704102), the National Natural Science Foundation of China (11674374, 11804378, 11888101 and 51672307), and the Key Research Program of Frontier Sciences, CAS (QYZDB-SSW-SLH008 and QYZDB-SSW-JSC035). This work has benefited from the bilateral collaboration F.R.S.-FNRS/NSFC (V4/345-DeM-229).

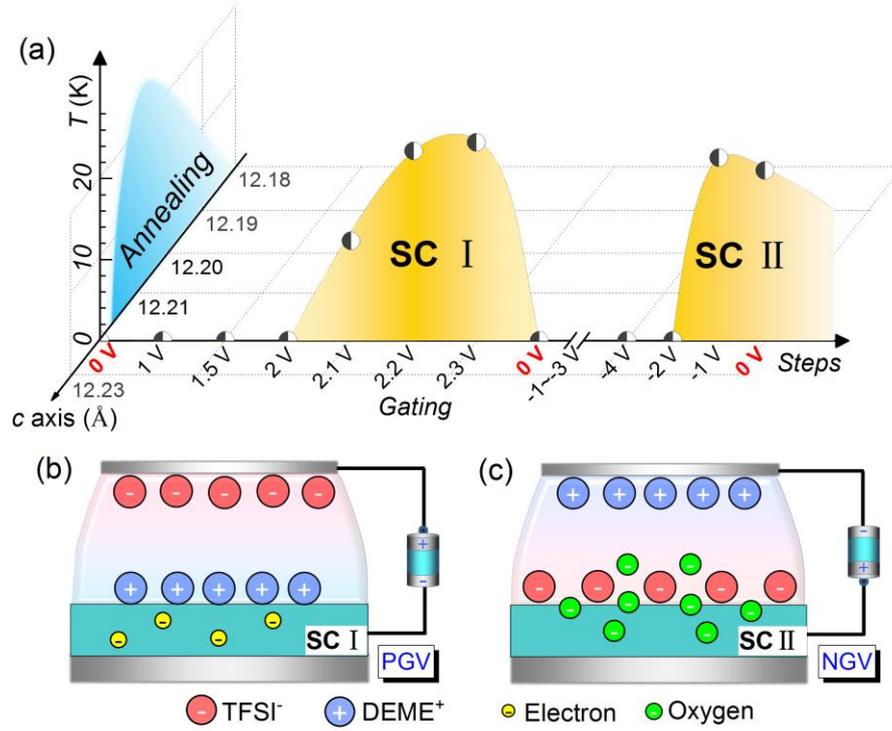

**Figure 1.** a) The dependence of the superconducting transition temperature $T_c$ as a function of *c*-axis lattice constant (a blue dome) and gate voltages (yellow domes, **SC I** and **SC II**). These different domes (blue and yellow) have been obtained with aid of annealing and gating methods. b, c) The schematic of the electron carrier doping and oxygen ions transfer induced with the help of ILG. (b) shows the configuration of the positive field gating by which the SC I dome arises and (c) shows negative field gating where the electrochemical reaction may arise leading to the **SC II** dome.

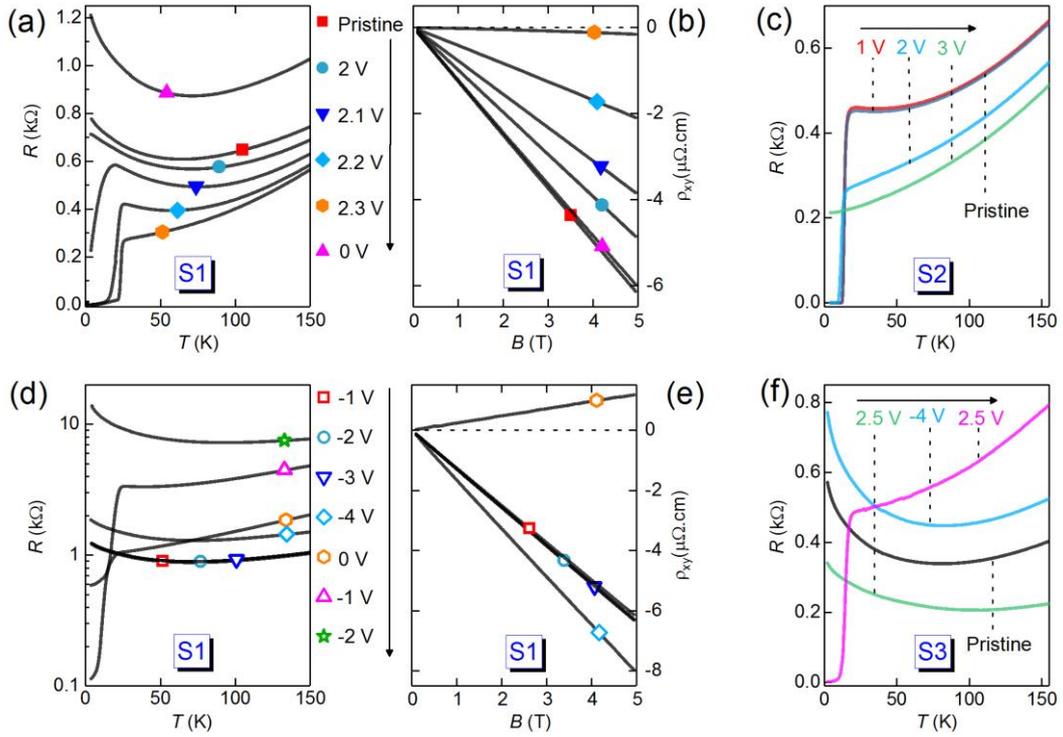

**Figure 2.** Electrical transport properties of $Pr_2CuO_{4\pm\delta}$ thin films. a, d) Temperature dependence of resistance for sample S1 at various gate voltages, with gating sequences noted by the arrows. b, e) Hall resistivity versus magnetic field at 30 K corresponding to the gating states of (a) and (d), respectively. c, f) Resistance as a function of temperature at different gate voltages for samples S2 and S3.

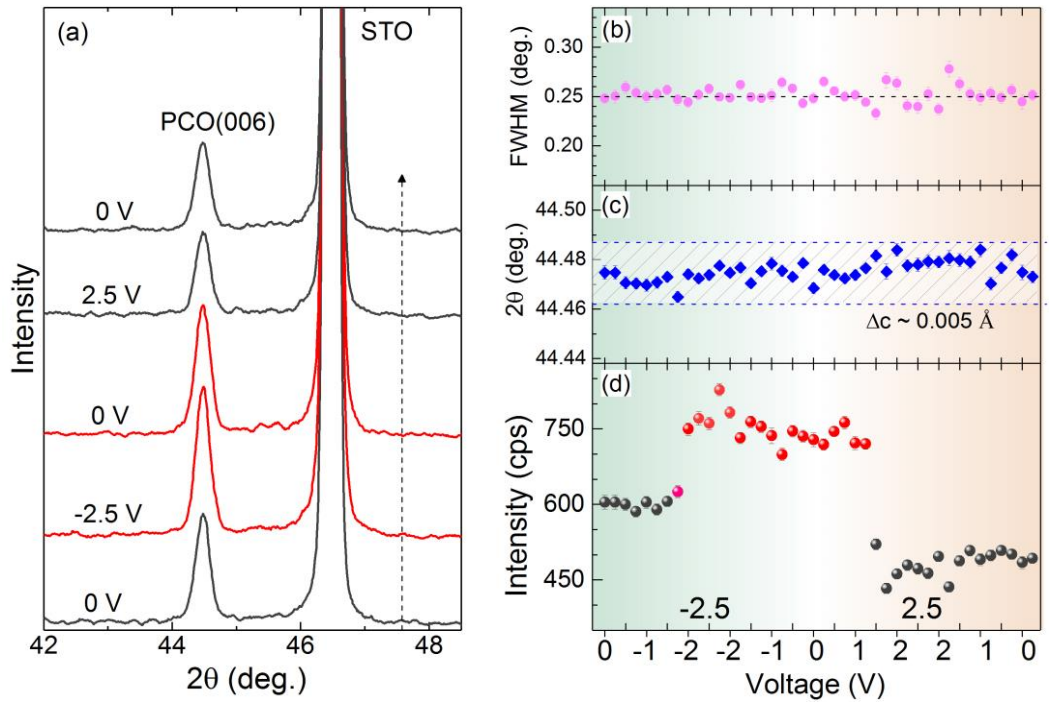

**Figure 3.** a) *In-situ* XRD measurements of $Pr_2CuO_{4\pm\delta}$ thin film at representative gate voltages, where the dotted arrow represents the sequence of ILG. b-d) Gate voltage dependence of full width at half maxima (FWHM) (b), angle (c) and amplitude (d) of $Pr_2CuO_{4\pm\delta}$ (006) peak.

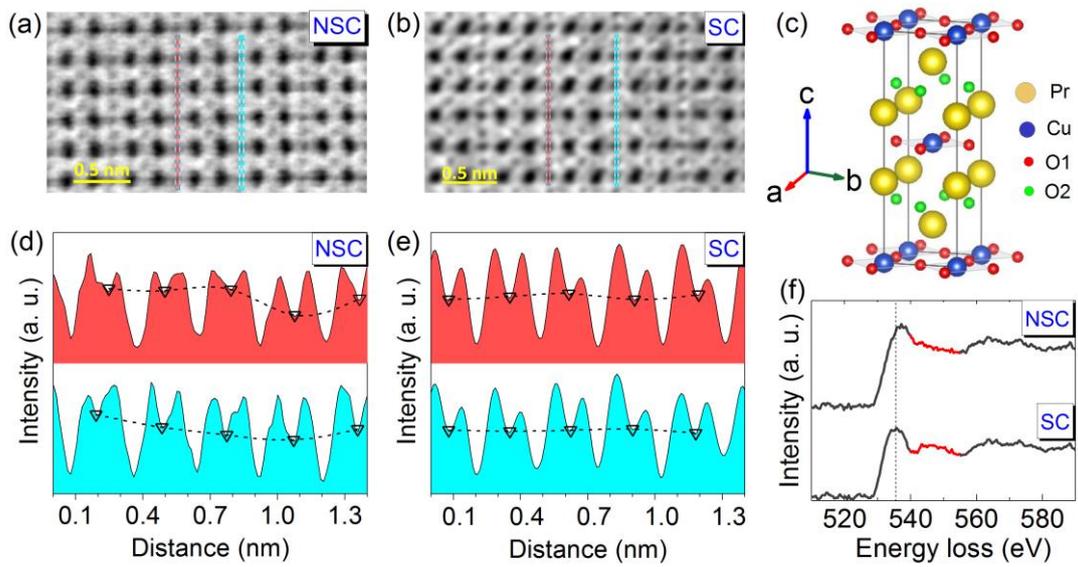

**Figure 4.** STEM and EELS data of $Pr_2CuO_{4\pm\delta}$ samples. a, b) Angular bright field images of NSC and SC samples, along the [110] zone axis. c) Crystal structure of $Pr_2CuO_{4\pm\delta}$. d, e) The line profiles of $CuO_2$ layers in the corresponding atomic column marked by colors in angular bright field images, where the deeper valley signifies higher occupancy of oxygen or copper. The valleys marked with inverted triangle symbols represent oxygen contrast. f) EELS profiles of SC and NSC samples normalized to the main peak of oxygen K edge. The curve difference between SC and NSC samples is highlighted by red color.